\documentstyle[mmatext,12pt]{article}
\newcommand{\bbb}{$B^0 \bar B^0$\ }
\newcommand{\e}{\mbox{e}}
\def\greaterthansquiggle{\raise.3ex\hbox%
                        {$>$\kern-.75em\lower1ex\hbox{$\sim$}}}
\def\lessthansquiggle{\raise.3ex\hbox{$<$\kern-.75em\lower1ex\hbox{$\sim$}}}
\newcommand{\be}{\begin{equation}}
\newcommand{\ee}{\end{equation}}
\newcommand{\ba}{\begin{eqnarray}}
\newcommand{\ea}{\end{eqnarray}}
\newcommand{\no}{\nonumber}

\newcommand{\lets}{\,\lessthansquiggle\,}
\normalsize
\begin{document}
\bibliographystyle{plain}
\begin{titlepage}
\begin{flushright}
UWThPh-1996-57\\
October 3, 1996\\
\end{flushright}
\vspace{15mm}
\begin{center}
{\Large \bf Quantum-Mechanical Interference over\\ Macroscopic Distances
in the $B^0 \bar B^0$ System}\\[70pt]

R.A. Bertlmann and W. Grimus\\
Institut f\"ur Theoretische Physik\\
Universit\"at Wien\\
Boltzmanngasse 5\\
A-1090 Vienna, Austria

\vspace{4cm}

{\bf Abstract}\\[7pt]
\end{center}
We argue that the \bbb state generated in the decay of
$\Upsilon(4S)$ is well suited for performing tests of
Einstein--Podolsky--Rosen correlations, i.e.,
quantum-mechanical interference effects over macroscopic
distances. Using measurements of the ratio $R=\,$(\#
like-sign dilepton events)/(\# opposite-sign dilepton events)
and of the $B_H$--$B_L$ mass difference we show that already presently
existing data strongly favour the contribution
of the interference term to $R$, as it is required by the rules
of quantum mechanics.\\[14pt]

\end{titlepage}

\section{Introduction}
Tests of quantum mechanics are of increasing interest in recent
years, in particular, the optical tests of quantum mechanics
carried out on systems of two correlated photons. Such 
systems---showing Einstein--Podolsky--Rosen (EPR) correlations---are
suitable to discriminate between quantum mechanics and any local
realistic (hidden variable) theory via Bell inequalities
\cite{Bell} (see, e.g., Ref. \cite{Bertlmann} for a quick introduction
into the field). All recent experiments using laser beams
confirm quantum mechanics in an impressive way (see, e.g.,
Refs. \cite{Aspect, Kwiat}) and they teach us that under
certain circumstances quantum systems extend over macroscopic scales.

We find it interesting and desireable to perform tests of
EPR correlations also with massive particles. Analogously to the
entangled photons one can create at $B$ factories states of
EPR-correlated \bbb pairs  as decay products of the upsilon
$\Upsilon(4S)$ resonance (see, e.g., Refs. \cite{Datta,Kayser}). 
More precisely, 
$B^0_d \bar B^0_d$ pairs are produced since $\Upsilon(4S)$ is not
heavy enough to decay into $B^0_s \bar B^0_s$. We drop the index
$d$ for convenience.

$B$ mesons have a lifetime of the order of a picosecond. If a \bbb
pair is produced by the decay of $\Upsilon(4S)$  there is very
little kinetic energy left per $B$ meson, namely roughly 10 MeV.
Multiplying the corresponding velocity $v$ of such a $B$ meson by its
lifetime one obtains $v \tau_{B^0} \approx 3 \times 10^{-2}$ mm.
This shows that in average the separation of the decaying $B$ mesons
originating in $\Upsilon(4S)$ is macroscopic. 
The \bbb system as the decay product of
$\Upsilon(4S)$ is a superposition of two states 
because the \bbb state inherits the charge
conjugation quantum number $C=-1$ of the $\Upsilon(4S)$. This
system offers therefore the possibility to
test, within particle physics, quantum-mechanical interference
over macroscopic distances.
Similar tests involving two-kaon systems have been proposed in
the past in Refs. \cite{Six,Selleri} and recently for
Da$\Phi$ne in Ref. \cite{ebe}.

To realize the above idea we consider the ratio $R=\,$(\#
like-sign dilepton events)/ 
(\#~opposite sign-dilepton events) of lepton pairs
generated in the decay chain $\Upsilon(4S) \rightarrow B^0 \bar
B^0 \rightarrow \ell^\pm \ell^\pm +$ anything. In order to
discriminate between quantum mechanics and local realistic
theories we introduce a {\it decoherence parameter} $\zeta$
\cite{ebe} such that the interference
term present in the quantum-mechanical calculation of $R$ is
multiplied by a factor $1~-~\zeta$ where $\zeta$ parameterizes
deviations from quantum mechanics. $\zeta$
is called decoherence parameter because at $\zeta = 1$
the interference is totally gone.
It turns out that, including this modification,
$R$ is a function of $\Delta m/\Gamma$, $\Delta \Gamma/2 \Gamma$ and
$\zeta$ with $\Delta m$, $\Delta \Gamma$ and $\Gamma$ being mass
difference, decay width difference and average decay width,
respectively, of the heavy and light neutral $B$ mass
eigenstates. There is also a parameter involved characterizing
CP violation in \bbb mixing. We will argue below that this
parameter can be set equal to one (no CP violation) and that
taking $\Delta \Gamma = 0$ is sufficient for our purpose. The
main idea of this paper is to compare the experimental value
$R_{\mbox{\scriptsize exp}}$ of $R$ measured at $\Upsilon(4S)$
with the theoretical expression $R(\Delta m/\Gamma, \zeta)$.
Taking $\Delta m$ from independent experiments which study the
time dependence of \bbb mixing and thus interference effects of
{\it single} $B$ states, the relation $R_{\mbox{\scriptsize exp}}=
R(\Delta m/\Gamma, \zeta)$ allows us to obtain information on
$\zeta$ and thus to test the long-range interference effects of
quantum mechanics in the \bbb system.

\section{The \bbb system}
To begin with we discuss the quantum mechanics of the \bbb system.
Phenomenologically, there are the two independent amplitudes 
\be
{\cal A}(B^0 \rightarrow f) \equiv A 
\quad \mbox{and} \quad
{\cal A}(\bar B^0 \rightarrow f) \equiv B
\ee
which enter into the description of the decays of the neutral $B$
mesons into an arbitrary final state $f$.
The mass eigenstates of the neutral $B$ mesons are given by
\ba
| B_H \rangle & = & p |B^0 \rangle + q |\bar B^0 \rangle\, ,
\no \\
| B_L \rangle & = & p |B^0 \rangle - q |\bar B^0 \rangle\, ,
\ea
with $ |p|^2 + |q|^2 = 1 $ and
\be \label{q/p}
\frac{q}{p} =
\frac{\Delta m - \frac{i}{2} \Delta \Gamma}{2 (M_{12}
- \frac{i}{2} \Gamma_{12})} =
\frac{2 (M_{12}^* - \frac{i}{2} \Gamma_{12}^*)}{\Delta m
- \frac{i}{2} \Delta \Gamma} =
\sqrt{\frac{
M_{12}^* - \frac{i}{2} \Gamma_{12}^*
}{
M_{12} - \frac{i}{2} \Gamma_{12}}}\, ,
\ee
where $\Delta m = m_H - m_L > 0$
($H$=heavy,
$L$=light),
$\Delta \Gamma = \Gamma_H - \Gamma_L$
and $M_{12} - \frac{i}{2} \Gamma_{12}$
is the off-diagonal matrix element in the effective time evolution
in the \bbb space \cite{gel}.
The positivity of $\Delta m$ fixes the sign of 
the square root in Eq.~(\ref{q/p}). The \bbb pair produced in
the decay of $\Upsilon(4S)$ is in the state
\be \label{psi}
\Psi(t=0) = \frac{1}{\sqrt{2}} \left( |B^0 \rangle \otimes |\bar B^0 \rangle
-|\bar B^0 \rangle \otimes |B^0 \rangle \right)
\ee
with charge conjugation quantum number $C=-1$ because the
$\Upsilon(4S)$ has quantum numbers $J^{CP}=1^{--}$ and its 
decay into \bbb proceeds via
strong interactions. The subsequent time evolution of
(\ref{psi}) is given by 
\ba
|B^0 (t) \rangle & = & g_+(t) |B^0 \rangle + \frac{q}{p} g_-(t)
|\bar B^0 \rangle \, ,
\nonumber\\
|\bar B^0 (t) \rangle & = & \frac{p}{q} g_-(t) |B^0 \rangle +
g_+(t) |\bar B^0 \rangle 
\ea
with
\be
g_\pm (t) = \frac{1}{2} \e^{-i(m - \frac{i}{2}\Gamma)t}
\left[
\e^{-\frac{i}{2}(\Delta m - \frac{i}{2}\Delta \Gamma)t} \pm
\e^{\frac{i}{2}(\Delta m - \frac{i}{2}\Delta \Gamma)t}
\right]
\ee
and
\be
m = \frac{1}{2} (m_H + m_L)\, , \;
\Gamma = \frac{1}{2} (\Gamma_H + \Gamma_L)\, .
\ee

After having introduced the basic formalism we now come to the
point where we modify the result of ordinary quantum mechanics
and subject this modification to a comparison with experimental
results. The class of observables we are interested in is the probability 
that $\Psi$ decays into final states $f_1$ and $f_2$ with
momenta $\vec p$ and $-\vec p$, respectively, in its restframe.
This probability 
is calculated by the integral \cite{car} 
\ba 
\lefteqn{N(f_1, f_2) =} \no \\ 
 & & \frac{1}{2} \int_0^\infty dt \int_0^\infty dt' \left\{
\left| \langle f_1 | B^0(t) \rangle  \right|^2 
\left| \langle f_2 | \bar B^0(t') \rangle  \right|^2 +
\left| \langle f_1 | \bar B^0(t) \rangle  \right|^2 
\left| \langle f_2 | B^0(t') \rangle  \right|^2  -  \right. \no \\
 & & -2 \, (1-\zeta) \, \mbox{Re} \, \Big[
\langle f_1 | B^0(t) \rangle^*
\langle f_2 | \bar B^0(t') \rangle^* 
\langle f_1 | \bar B^0(t) \rangle 
\langle f_2 | B^0(t') \rangle \Big] \bigg\}.  \label{N12}
\ea
The last term in Eq. (\ref{N12}) is the usual
quantum-mechanical interference term as it results from the two
summands of the wave function (\ref{psi}) modified by a
factor $1-\zeta$ \cite{ebe}. In the following we will rather
arbitrarily assume that $0 \le \zeta \le 1$ to incorporate
quantum mechanics with $\zeta = 0$ at one end of the interval and
no interference corresponding to $\zeta = 1$ at the other end.
Our aim is to test which range of $\zeta$ is experimentally
allowed if we use information on semileptonic decays of the \bbb
system. To apply Eq. (\ref{N12}) we have to perform the
integrals and we arrive at the general formula
\ba \label{N}
\lefteqn{N(f_1, f_2) =} \no \\ 
 & & \frac{1}{2} \left\{ I_1 \bigg|A_1B_2-B_1A_2\bigg|^2 + 
     I_2 \left|\frac{p}{q}A_1A_2-\frac{q}{p}B_1B_2\right|^2 \right\} + \no \\
 & & + \zeta \, \mbox{Re} \, \left\{ \left( 
       I_+ A_1^*B_1 + I_- \left(\frac{q}{p}\right)^* \frac{p}{q} B_1^*A_1 +
       I_{+-} \frac{p}{q} |A_1|^2 + I_{-+} \left(\frac{q}{p}\right)^*
       |B_1|^2 \right) \right. \no \\
 & & \left. \cdot \left( 
       I_+ B_2^*A_2 + I_- \left(\frac{p}{q}\right)^* \frac{q}{p} A_2^*B_2 +
       I_{+-} \frac{q}{p} |B_2|^2 + I_{-+} \left(\frac{p}{q}\right)^*
       |A_2|^2 \right) \right\}
\ea
with
\ba
I_1 & = & \int_0^\infty dt \int_0^\infty dt' \, 
| g_+(t)g_+(t') - g_-(t)g_-(t') |^2 = 
I_+^2 + I_-^2 - 2 \, \mbox{Re} \, (I_{+-})^2 = \frac{1}{\Gamma}
I_+\, ,\no \\ 
I_2 & = & \int_0^\infty dt \int_0^\infty dt' \,
| g_+(t)g_-(t') - g_-(t)g_+(t') |^2 = 
2 I_+ I_- - 2 | I_{+-} |^2 = \frac{1}{\Gamma} I_- \, ,
\ea
\ba
I_\pm & = & \int_0^\infty dt \, | g_\pm(t) |^2 = 
\frac{1}{2\Gamma} \left( \frac{1}{1-y^2} \pm \frac{1}{1+x^2}
\right) \, ,\no \\ 
I_{+-} & = & \int_0^\infty dt \, g_+(t)^* g_-(t) =
-\frac{1}{2\Gamma} \left( \frac{y}{1-y^2} + i\frac{x}{1+x^2}
\right) \, ,
\ea
and $x$ and $y$ are defined as
\be
x = \frac{\Delta m}{\Gamma} \quad \mbox{and} \quad 
y = \frac{\Delta \Gamma}{2 \Gamma}.
\ee
Furthermore, the relation $I_{-+}=(I_{+-})^*$ is valid.

In principle, measurements of $N(f_1, f_2)$ for any $f_1$, $f_2$
could be used to
obtain information on $x$ (see Ref. \cite{ARGUS0}), $y$ and $\zeta$. 
In this case one would have to
know the quantities $|A_1B_2-B_1A_2|$, 
$|\frac{p}{q}A_1A_2-\frac{q}{p}B_1B_2|$, etc. which, in general,
require additional experimental information. However, for semileptonic
decays the situation is very simple because in lowest order in
weak interactions only the tree-level $W$ exchange graphs are
responsible for such decays. In addition, since the quark
content of \bbb is given by $B^0 = (\bar b d)$ and $\bar B^0 =
(b \bar d)$ the lepton $\ell^+$ in the final state tags $B^0$
whereas $\ell^-$ tags $\bar B^0$. Therefore,
with $f_+ \equiv X \ell^+
\nu_\ell$ and $f_- \equiv \bar X \ell^- \bar \nu_\ell$ and the labels
$+$, $-$ pertaining to $f_+$, $f_-$, respectively, we have
\be \label{semileptamp}
|A_+| = |B_-| \quad \mbox{and} \quad B_+ = A_- = 0.
\ee
In these final states $X$ denotes an arbitrary kinematically
allowed hadronic state and $\bar X$ its charge-conjugate counterpart.
Defining $N_{++} \equiv N(f_+, f_+)$, etc., and using Eq.
(\ref{semileptamp}) we obtain the following very simple
expression for $N(f_1, f_2)$, Eq. (\ref{N}),
in the case of semileptonic decays:
\ba \label{N++}
N_{++} & = &
\frac{1}{2} |A_+|^4 \left| \frac{p}{q} \right|^2 ( I_2 + 2 \zeta
|I_{+-}|^2 ), \\ 
\label{N--}
N_{--} & = &
\frac{1}{2} |B_-|^4 \left| \frac{q}{p} \right|^2 ( I_2 + 2 \zeta
|I_{+-}|^2 ), \\
\label{N+-}
N_{+-} = N_{-+} & = &
\frac{1}{2} |A_+|^2 |B_-|^2 ( I_1 + 2 \zeta \, \mbox{Re} \, (I_{+-})^2 ).
\ea
Defining the ratio of like-sign dilepton events to opposite-sign
dilepton events \cite{pai,oku}
\be
R \equiv \frac{N_{++} + N_{--}}{N_{+-} + N_{-+}}
\ee
the amplitudes cancel and we find $R$ as a function of $|p/q|$, $x$, $y$
and $\zeta$:
\be \label{R}
R = \frac{1}{2} \left( \left| \frac{p}{q} \right|^2 + 
                       \left| \frac{q}{p} \right|^2  \right)
    \frac{x^2 + y^2 + \zeta \left[ y^2 \frac{1+x^2}{1-y^2} + 
                                   x^2 \frac{1-y^2}{1+x^2} \right] }%
         {2+x^2-y^2 + \zeta \left[ y^2 \frac{1+x^2}{1-y^2} -
                                   x^2 \frac{1-y^2}{1+x^2} \right]}.
\ee

It is well known that a deviation of $|p/q|$ from 1 is a signal
for CP violation in \bbb mixing. A suitable measure for $|p/q|$ and
CP violation in mixing is thus given by \cite{oku}
\be \label{CP}
A_{\mbox{\scriptsize CP}} \equiv \frac{N_{++} - N_{--}}{N_{++} + N_{--}} =
\frac{|\frac{p}{q}|^2 - |\frac{q}{p}|^2}{|\frac{p}{q}|^2 + |\frac{q}{p}|^2}.
\ee
To derive this formula, Eqs. (\ref{N++}), (\ref{N--}) and
(\ref{N+-}) have been used which correspond to odd relative
angular momentum of the \bbb pair. It is easy to show with the
methods expounded here that the same formula (\ref{CP}) is
valid for even relative angular momentum. Moreover, Eq.
(\ref{CP}) is also valid for any statistical mixture of odd and
even \cite{gri} and does not depend on the parameter $\zeta$ which could
even be different for odd and even. This shows that it is consistent 
to take any measurement of $A_{\mbox{\scriptsize CP}}$ and use it
as information on $|p/q|$ in $R$ (\ref{R}). A recent measurement
of the CDF Collaboration \cite{CDF} gives $A_{\mbox{\scriptsize CP}} = (2.4
\pm 6.3 \,(\mbox{stat}) \pm 3.3 \,(\mbox{sys})) \times 10^{-2}$.
The factor in front of $R$ which depends on $|p/q|$ is expressed
by $A_{\mbox{\scriptsize CP}}$ as
\be \label{factor}
\frac{1}{2} \left( \left| \frac{p}{q} \right|^2 + 
                       \left| \frac{q}{p} \right|^2  \right) =
(1-A_{\mbox{\scriptsize CP}}^2)^{-1/2} \approx 
1 + \frac{1}{2} A_{\mbox{\scriptsize CP}}^2.
\ee
With the above value of $A_{\mbox{\scriptsize CP}}$ the quantity
(\ref{factor}) differs less than a percent from 1. In view of
the experimental errors associated with $R$ and $x$ we will
simply set (\ref{factor}) equal to 1 in the rest of this paper.

\section{Discussion of the experimental data}
Having disposed of $|p/q|$, there remain three variables in $R$,
namely $x=\Delta m/\Gamma$, $y$ and $\zeta$. To test the
quantum-mechanical interference term, i.e., to get information
on $\zeta$, we want to take $x$ from
measurements of the time dependence of \bbb mixing 
\cite{ALEPH,DELPHI,L3,OPAL} and compare $R$ with $R_{\mbox{\scriptsize exp}}$ 
measured at the
$\Upsilon (4S)$ \cite{ARGUS,CLEO}. In the concrete, we apply the
following procedure. We take the values of $\Delta m$ from the
results of the LEP experiments ALEPH \cite{ALEPH}, DELPHI \cite{DELPHI}, L3
\cite{L3} and OPAL \cite{OPAL} 
which are $\Delta m = 0.436 \pm 0.033 \:\hbar/$ps, 
$\Delta m = 0.531
\begin{array}{l} +0.050 \\ -0.046 \end{array} \pm 0.078
\:\hbar$/ps, 
$\Delta m = 0.496 \begin{array}{l} +0.055\\ -0.051 \end{array}
\pm 0.043 \:\hbar$/ps and 
$\Delta m = 0.548 \pm 0.050 
\begin{array}{l} +0.023 \\ -0.019 \end{array}\hbar/$ps,
respectively. The first error is the statistical and the second one
the systematic. For each experiment, we simply add 
the squares of the statistical and systematic error (we
select the larger value where positive and negative errors are
different) and
use the law of combination of errors to get the combined
value of $\Delta m$. After division by $\tau_{B^0} = (1.56 \pm
0.06)$ ps \cite{RPP} we arrive at the final value $\bar x = 0.74
\pm 0.05$ which will
be used in the figures. As for $R$ we take the experimental input
$R_{\mbox{\scriptsize exp}} = 0.194 \pm 0.062 \pm 0.054$
obtained by ARGUS \cite{ARGUS} and $R_{\mbox{\scriptsize exp}} =
0.187 \pm 0.022 \pm 0.025 \begin{array}{l} +0.040 \\ -0.030
\end{array}$, the result of the CLEO Collaboration \cite{CLEO},
where the third error reflects a $\pm 15$ \% uncertainty in the
assumption that charged and neutral $B$ pairs contribute equally
to dilepton events. 
Performing the same steps as for $\Delta m$ we
obtain $\bar R_{\mbox{\scriptsize exp}} = 0.189 \pm 0.044$.

It remains to discuss $y$ in the context of the determination of
the decoherence parameter. The Standard Model predicts a
very small difference between the lifetimes of the heavy and the
light neutral $B$ meson such that $|y|/x \lets 10^{-2}$ (see,
e.g., Ref. \cite{y}). This alone would already constitute a
strong motivation for putting $y=0$ in $R$ (\ref{R}).
Furthermore, plotting $R$ as a function of the decoherence
parameter $\zeta$ $(0 \le \zeta
\le 1)$ and comparing the curves with $y=0$ and $y=0.1$ there is
practically no difference. Last but not least, studying the
ratio $R$ as a function of $y$ and $\zeta$ 
numerically reveals that with increasing
$y^2$ the restriction on $\zeta$ gets stronger. Therefore, for our
purpose of getting information on the quantum-mechanical
interference term it is sufficient to study $R$ as a function of
$\zeta$ with $y=0$.

This is done in Figs. 1 and 2. The three curves in Fig. 1
correspond to $R$ with $y=0$ and the three $x$ values $\bar x -
\Delta \bar x$ (lower curve), $\bar x$ (middle curve) and $\bar
x + \Delta \bar x$ (upper curve). The horizontal lines indicate 
the mean value
$\bar R_{\mbox{\scriptsize exp}}$ and $\bar R_{\mbox{\scriptsize
exp}} \pm \Delta \bar R_{\mbox{\scriptsize exp}}$. In Fig. 2 we
have again plotted $R$ and $R_{\mbox{\scriptsize exp}}$ but
the error bands correspond to 1.64 standard deviations
or 90 \% CL if the distributions are Gaussian. 

As a side-remark we want to stress that the method discussed
here can also be used to get a bound on $|y|$. For simplicity we
assume quantum mechanics to be valid ($\zeta = 0$) and compare
$R$ as a function of $y$ with $R_{\mbox{\scriptsize exp}}$.
Then we obtain $|y| \le 0.40$ at 90 \% CL. This 
shows that $y = \Delta \Gamma /2 \Gamma$ and
thus the difference in the decay widths of the heavy and light
neutral $B$ mesons is only mildly restricted by present data.
There is still a large gap between experimental information and
the Standard Model prediction for $y$.

\section{Conclusions}
We observe that the overlap of the allowed areas of $R$ and 
$R_{\mbox{\scriptsize exp}}$ restricts the decoherence
parameter to $\zeta \le 0.26$ in Fig. 1 and $\zeta \le 0.53$ in
Fig. 2. This result conformes nicely with quantum mechanics and
leaves little room for local realistic theories ($\zeta = 1$).
Of course, our statistical analysis
is rather crude and the experimental errors of $x$ and 
$R_{\mbox{\scriptsize exp}}$ are
large. Nevertheless, there is a clear sign of long-range 
interference effects in
\bbb in agreement with quantum mechanics.  This
is not so surprising in view of the overwhelming success of
quantum mechanics. We expect that with the
improvement of the experimental errors the bound on $\zeta$ will
become much tighter in the future. However, we also notice
that the mean value of $R$ at $\zeta=0$ is slightly
higher than the mean value of 
$R_{\mbox{\scriptsize exp}}$. This acts in favour of smaller
bounds on $\zeta$ (see figures) and has to be kept 
in mind when considering their above numerical values.
Adding a note of scepticism
concerning tests like the one discussed here, we want to remark
that changing quantum mechanics in one point, in the present
case in the two-particle interference term, but assuming its
validity in all other domains, e.g., one-particle interference
terms from which $\Delta m$ is extracted, is an arbitrary
procedure. However, since no consistent {\it local} theory encompassing
quantum mechanics is known, all parameterizations of deviations
from quantum mechanics involve a certain amount of arbitrariness.

\newpage

\begin{figure}
\centerline{\input{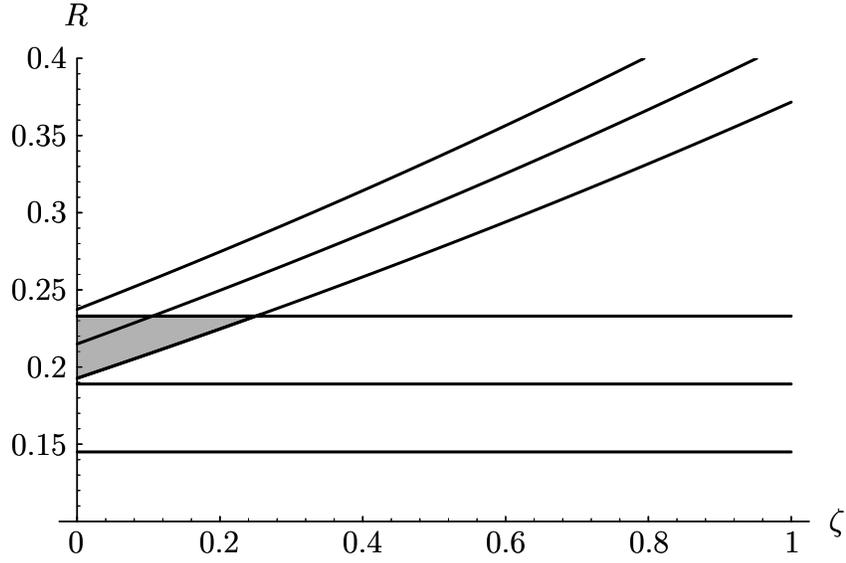}}
\caption{$R$ as a function of the decoherence parameter $\zeta$
for $x= \bar x, \; \bar x \pm \Delta \bar x$. The horizontal
lines indicate the mean value $\bar R_{\mbox{\protect\scriptsize exp}}$
and $\bar R_{\mbox{\protect\scriptsize exp}} \pm \Delta 
\bar R_{\mbox{\protect\scriptsize exp}}$ of the experimental measurement
of $R$. The shaded region shows the overlap of the two error bands.}
\end{figure}

\begin{figure}
\centerline{\input{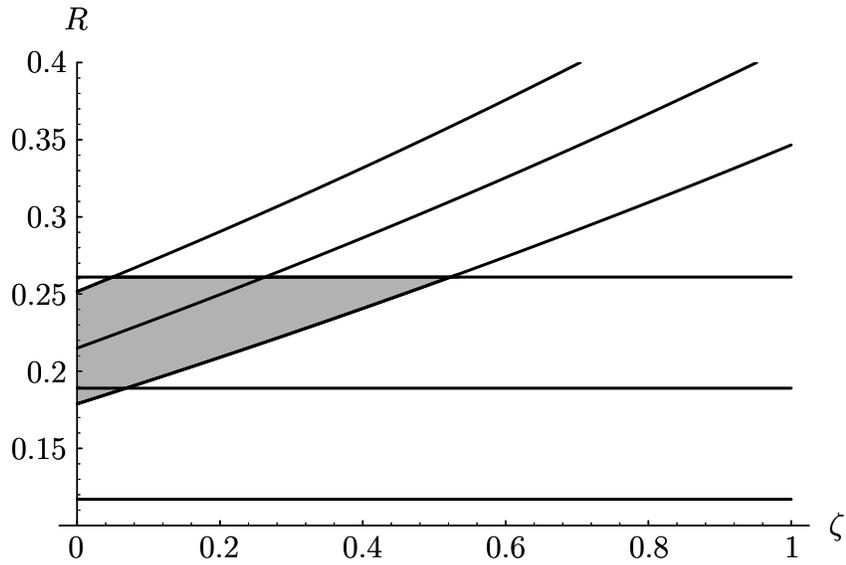}}
\caption{The same as in Fig. 1 but the error bands correspond to
1.64 $\sigma$ or 90 \% CL.}
\end{figure}

\end{document}